\documentclass[cameraready]{Interspeech}

\usepackage{multirow}

\title{Stabilizing Instruction Supervision for Instruct-TTS via Controllable Diversification and Drift Filtering}

\author[affiliation={1}]{Yizhong}{Geng}
\author[affiliation={1}]{Kecan}{Mao}
\author[affiliation={1}]{Qifei}{Li}
\author[affiliation={1}]{Cong}{Wang}
\author[affiliation={1}]{Yingming}{Gao}
\renewcommand{\authorsep}{, \\}
\author[affiliation={2}]{Ruimin}{Wang}
\author[affiliation={2}]{Chunfeng}{Wang}
\author[affiliation={2}]{Hao}{Li}
\author[affiliation={1,*}]{Ya}{Li}

\address{
$^1$ Beijing University of Posts and Telecommunications, China \\
$^2$ Li Auto, China
}

\email{yli01@bupt.edu.cn}


\makeatletter
\renewcommand{\paragraph}[1]{\vspace{0.5ex}\noindent\textbf{#1}}
\makeatother

\begin{document}

\maketitle

\hypersetup{
  pdftitle={Stabilizing Instruction Supervision for Instruct-TTS via Controllable Diversification and Drift Filtering},
  pdfauthor={Yizhong Geng; Kecan Mao; Qifei Li; Cong Wang; Yingming Gao; Ruimin Wang; Chunfeng Wang; Hao Li; Ya Li},
  pdfkeywords={Instruct-TTS, label-to-instruction, instruction instability, drift taxonomy, data-centric stabilization}
}

\begingroup
\renewcommand\thefootnote{}
\footnotetext{$^*$ Corresponding authors.}
\endgroup

\begin{abstract}
Instruct-TTS systems expand structured style labels into natural-language training instructions through LLM rewriting, yet we find that over 40\% of unconstrained rewrites contain semantic drift that corrupts supervision and weakens generalization. We formalize this problem as instruction supervision instability and propose a data-centric stabilization recipe that jointly improves coverage and fidelity through three mechanisms: controllable instruction diversification for systematic expansion, LLM-based drift filtering for quality control, and attribute-aligned supervision that grounds prosody control in acoustic perturbations. On the Chinese split of InstructTTSEval, our recipe raises instruction-following from 34.5\% without fine-tuning and 51.0\% with naive fine-tuning to 56.4\%, while constrained rewriting reduces drift from 40.4\% to 15.4\%. Ablations confirm the three mechanisms are complementary, and the drift taxonomy may generalize to instruction-driven generation beyond TTS. \footnote{Audio samples are available on the \href{https://piedpiperg.github.io/instruct-tts-stabilizer/\#audio-demos}{project demo page}.}
\end{abstract}

\section{Introduction}

Text-to-speech synthesis is shifting from fixed tags and numeric controls to Instruct-TTS, where users specify style, emotion, and prosody through free-form natural language \cite{guo2023prompttts, leng2023prompttts, yang2024instructtts, zhou2024voxinstruct}. While this interface broadens applicability in audiobooks, assistants, and content creation, it hinges on large-scale instruction--speech supervision with broad coverage and high semantic fidelity \cite{wang2023self}, which is difficult to collect at scale.

\begin{figure}[t]
  \centering
  \includegraphics[width=\columnwidth]{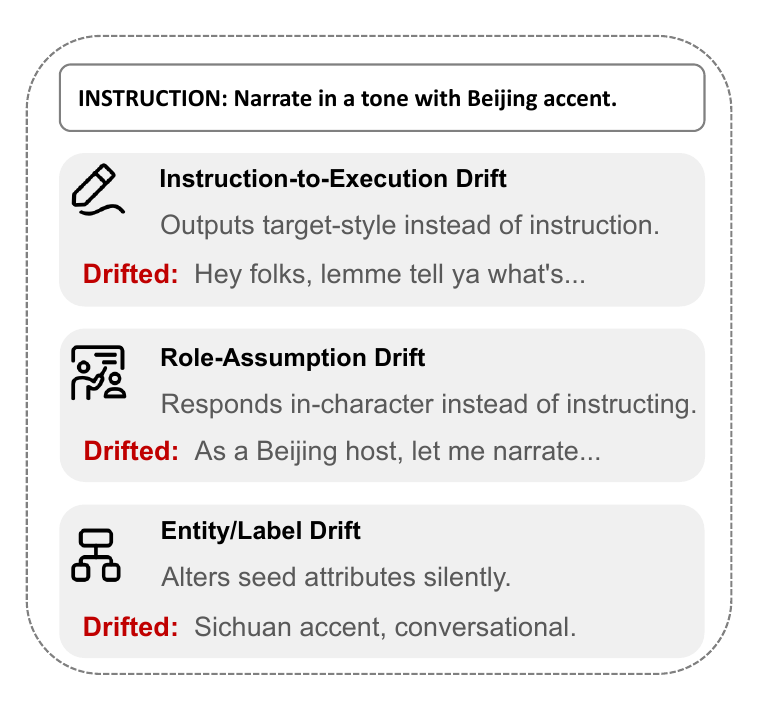}
  \caption{Three drift types in LLM-rewritten instructions: a drifted output versus the expected faithful instruction from the same seed attributes.}
  \label{fig:drift_examples}
\end{figure}

A common workaround is to convert structured style or attribute labels into natural-language instructions using templates or LLM rewriting \cite{guo2023prompttts, wang2023self, leng2023prompttts}, a process we refer to as the \emph{label-to-instruction} pipeline. In NLP, Self-Instruct and Evol-Instruct have shown that LLM-generated instructions can effectively train instruction-following models \cite{wang2023self, xu2023wizardlm}, and LLM-as-judge methods are widely adopted for data filtering \cite{zheng2023judging, wang2022self}. However, these pipelines largely assume that LLM output quality is manageable through standard prompting. In the TTS setting, the label-to-instruction pipeline suffers from two coupled weaknesses. First, generated prompts are homogeneous in phrasing and viewpoint, providing \emph{limited coverage} of real user expressions \cite{wang2023self, xu2023wizardlm}. Second, LLM rewrites are prone to \emph{semantic drift}: they silently change the intended control meaning \cite{xu2023wizardlm, fu2024learning, sun2023evaluating}, e.g., producing an execution-like utterance instead of a control instruction, adopting an in-character persona, or altering seed attributes (Fig.~\ref{fig:drift_examples}). These two weaknesses together corrupt training signals, which may explain why naive fine-tuning on such data can underperform a no-SFT baseline \cite{alajrami2025fine, zhang2025best}.

\begin{figure*}[t]
  \centering
  \includegraphics[width=\textwidth,trim=10 10 10 10,clip]{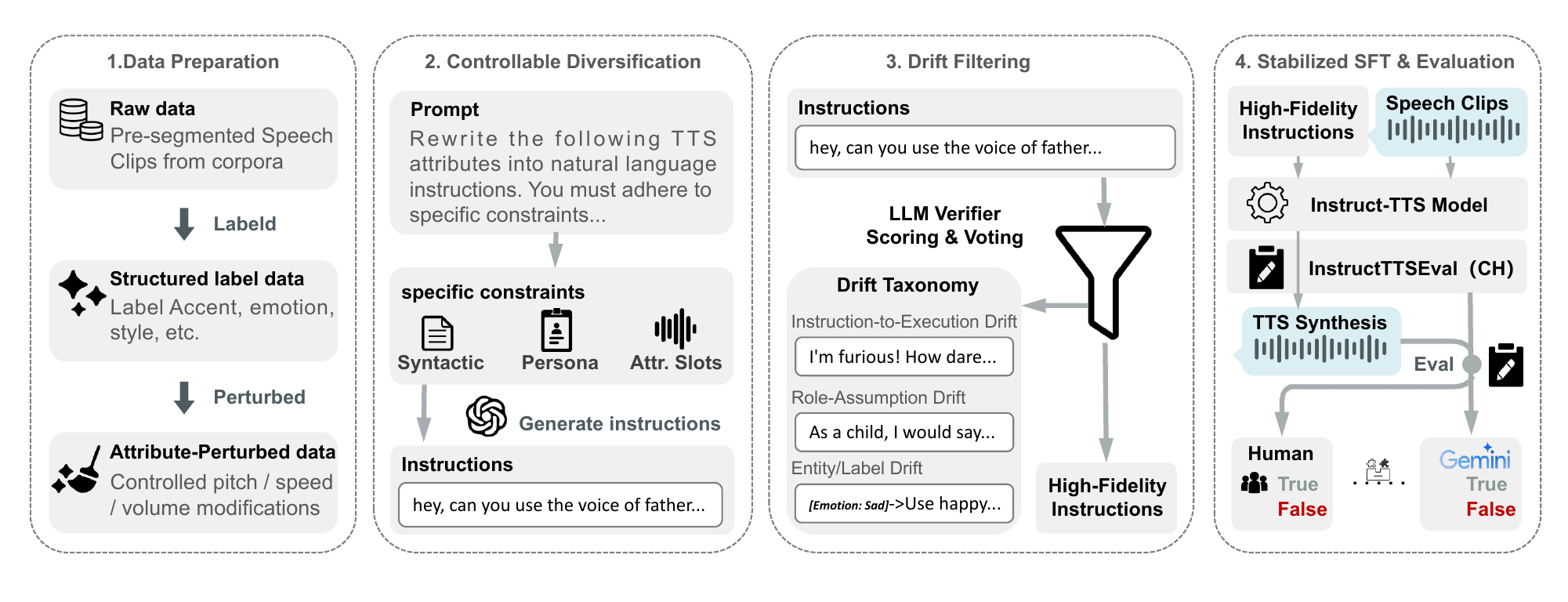}
  \caption{Overview of our data-centric stabilization recipe: (1)~data preparation with structured labeling and pitch/speed/volume perturbation; (2)~controllable instruction diversification with persona, syntactic pattern, and attribute slot constraints; (3)~drift filtering via LLM verifier scoring and voting; (4)~stabilized fine-tuning and evaluation on InstructTTSEval.}
  \label{fig:overview}
\end{figure*}

We unify the two weaknesses above under the concept of \emph{instruction supervision instability} and argue that it is fundamentally a data quality problem rather than a model architecture limitation: the model itself is capable of following instructions, but the training signal is too noisy and narrow to teach it reliably. This perspective suggests a \emph{data-centric} remedy: improving the instruction data along both coverage and fidelity dimensions simultaneously \cite{geng2026bridging}. To quantify the problem, we annotate LLM-expanded instructions and identify three recurring drift patterns---\emph{instruction-to-execution}, \emph{role-assumption}, and \emph{entity/label} drift---finding that over 40\% of unconstrained rewrites are supervision-corrupting. Crucially, expanding diversity alone does not ensure fidelity, and filtering alone does not expand coverage, so the two must be addressed jointly.

Guided by this analysis, we assemble a data-centric stabilization recipe (Fig.~\ref{fig:overview}) from three complementary mechanisms. \emph{Controllable instruction diversification} generates semantically equivalent variants by constraining persona, syntactic pattern, and attribute slots, yielding systematic coverage expansion without free-form drift. \emph{Drift filtering} removes residual supervision-corrupting rewrites via an LLM verifier with scoring and voting \cite{zheng2023judging, wang2022self}, guided by our drift taxonomy. \emph{Attribute-aligned supervision} pairs pitch/speed/volume perturbations with natural-language attribute prompts \cite{ko2015audio, jaitly2013vocal}, grounding low-level prosody controllability that label-to-instruction rewrites alone cannot reliably cover. On the Chinese split of InstructTTSEval \cite{huang2025instructttseval}, the full recipe consistently outperforms both a no-SFT baseline and naive fine-tuning on unfiltered rewrites. Our contributions are:

\begin{itemize}
  \item \textbf{Problem characterization.} We define \emph{instruction supervision instability} and quantify it with a three-category drift taxonomy, revealing that over 40\% of unconstrained LLM rewrites corrupt supervision.
  \item \textbf{Data-centric stabilization.} We combine controllable diversification, drift filtering, and attribute-aligned supervision into a unified recipe and show through ablation that the three mechanisms are complementary.
  \item \textbf{Empirical validation.} On InstructTTSEval (Chinese) \cite{huang2025instructttseval}, our recipe yields consistent gains in instruction-following accuracy and controllability, with drift filtering as the single most impactful component.
\end{itemize}

\section{Related Work}
\label{sec:related}

\paragraph{Instruct-TTS and instruction data quality.}
PromptTTS \cite{guo2023prompttts}, PromptTTS\,2 \cite{leng2023prompttts}, InstructTTS \cite{yang2024instructtts}, and VoxInstruct \cite{zhou2024voxinstruct} progressively extend natural-language control from speaking style to multi-dimensional instructions covering emotion, accent, and prosody, while ParlerTTS scales the paradigm with large annotated corpora \cite{lyth2024natural}. These efforts focus on model architectures and training objectives, treating the upstream instruction data as given. Our work is orthogonal: we target the coverage and fidelity of the label-to-instruction supervision itself, systematically characterizing domain-specific semantic drift that standard NLP instruction pipelines \cite{wang2023self, xu2023wizardlm, zheng2023judging} do not address.

\paragraph{Controllable speech synthesis.}
Conventional controllable TTS relies on reference audio embeddings or discrete style tags \cite{wang2018style, valle2020mellotron}, while low-level attributes such as pitch, speed, and volume are typically adjusted through signal processing or model-internal controls \cite{ren2020fastspeech, bai2025hq}. Our attribute-aligned supervision bridges this gap by pairing parameterized perturbations with natural-language prompts, providing high-fidelity grounded signal for dimensions that free-form rewrites cover with limited reliability.

\begin{table*}[t]
  \centering
  \caption{Main results on InstructTTSEval (Chinese). Acc.\ (\%) denotes Gemini-judged instruction-following accuracy across APS, DSD, and RP tasks. Avg.\ is the arithmetic mean of the three. CER (\%) is character error rate of synthesized speech. NMOS and CMOS are human-rated naturalness and controllability scores (1--5 scale, 95\% CI). Upper block: external baseline; lower block: internal recipe variants on the same CosyVoice\,2 backbone. Best in our systems in \textbf{bold}.}
  \label{tab:main}
  \setlength{\tabcolsep}{5pt}
  \renewcommand{\arraystretch}{1.15}
  \begin{tabular}{cl cccc c cc}
    \toprule
    & & \multicolumn{4}{c}{\textbf{Instruction Following Acc.\ (\%)}} & & \multicolumn{2}{c}{\textbf{Human Eval (MOS)}} \\
    \cmidrule(lr){3-6} \cmidrule(lr){8-9}
    & \textbf{Setting} & \textbf{APS} & \textbf{DSD} & \textbf{RP} & \textbf{Avg.} & \textbf{CER}$\downarrow$ & \textbf{NMOS}$\uparrow$ & \textbf{CMOS}$\uparrow$ \\
    \midrule
    Ext.
    & VoxInstruct        & 47.5 & 52.3 & 42.6 & 47.5 & 22.5 & 3.18{\scriptsize$\pm$.08} & 3.12{\scriptsize$\pm$.09} \\
    \midrule
    \multirow{3}{*}{Ours}
    & Base (no SFT)       & 20.2 & 45.5 & 37.7 & 34.5 & 35.0 & 3.30{\scriptsize$\pm$.09} & 3.17{\scriptsize$\pm$.09} \\
    & Naive SFT           & 45.2 & 62.2 & 45.5 & 51.0 & 19.2 & 3.46{\scriptsize$\pm$.08} & 3.22{\scriptsize$\pm$.09} \\
    & Full Recipe         & \textbf{49.5} & \textbf{65.4} & \textbf{54.4} & \textbf{56.4} & \textbf{17.7} & \textbf{4.16}{\scriptsize$\pm$.06} & \textbf{4.16}{\scriptsize$\pm$.06} \\
    \bottomrule
  \end{tabular}
\end{table*}

\begin{table}[t]
  \centering
  \caption{Drift rates (\%) of LLM-expanded instructions. \emph{Unconstrained}: free-form LLM rewriting; \emph{Constrained}: rewriting with persona, pattern, and slot controls (800 annotated samples).}
  \label{tab:drift}
  \setlength{\tabcolsep}{5pt}
  \renewcommand{\arraystretch}{0.95}
  \small
  \begin{tabular}{l cc}
    \toprule
    \textbf{Drift Type} & \textbf{Unconstrained} & \textbf{Constrained} \\
    \midrule
    Instruction-to-Execution & 18.3 & 7.1  \\
    Role-Assumption          & 12.7 & 4.5  \\
    Entity/Label             &  9.4 & 3.8  \\
    \midrule
    Total                    & 40.4 & 15.4 \\
    \bottomrule
  \end{tabular}
\end{table}

\begin{table}[t]
  \centering
  \caption{Ablation on InstructTTSEval (Chinese), removing each component from the Full Recipe. Metrics follow Table~\ref{tab:main}.}
  \label{tab:ablation}
  \setlength{\tabcolsep}{5pt}
  \renewcommand{\arraystretch}{0.95}
  \small
  \begin{tabular}{l cccc c}
    \toprule
    & \multicolumn{4}{c}{\textbf{Instruction Following Acc.\ (\%)}} & \\
    \cmidrule(lr){2-5}
    \textbf{Setting} & \textbf{APS} & \textbf{DSD} & \textbf{RP} & \textbf{Avg.} & \textbf{CER}$\downarrow$ \\
    \midrule
    Full Recipe          & 49.5 & 65.4 & 54.4 & 56.4 & 17.7 \\
    \; -- Ctrl.\ Div.   & 45.8 & 60.1 & 49.6 & 51.8 & 18.4 \\
    \; -- Drift Filt.   & 43.1 & 57.3 & 46.2 & 48.9 & 19.8 \\
    \; -- Attr.\ Sup.   & 42.7 & 63.8 & 52.7 & 53.1 & 18.1 \\
    \bottomrule
  \end{tabular}
\end{table}

\section{Method}

\subsection{Attribute-Aligned Supervision}
\label{sec:data_prep}

We start from a speech corpus pre-segmented into short clips, each annotated with structured seed attributes such as accent, emotion, gender, and speaking style (Fig.~\ref{fig:overview}, Stage~1).
Since these high-level labels alone provide limited coverage of low-level prosody controls, we additionally apply parameterized perturbations along pitch, speaking rate, and volume to each clip \cite{ko2015audio,jaitly2013vocal,ren2020fastspeech}.
Perturbation settings are mapped to concise natural-language attribute prompts via lightweight templates, yielding paired (perturbed speech, attribute prompt) examples that ground low-level controllability in high-fidelity acoustic manipulations rather than potentially drifted LLM outputs.

\subsection{Controllable Instruction Diversification}
\label{sec:diversification}

To expand the limited coverage of template-based instructions, we generate instruction variants from seed attributes $a$ under three explicit constraints (Fig.~\ref{fig:overview}, Stage~2):
\emph{persona} specifies the speaker viewpoint to diversify pragmatic intent;
\emph{syntactic pattern} constrains the surface form to mitigate template over-reliance;
and \emph{attribute slots} require all target attributes as non-droppable slots, preventing entity/label drift through omission.
These constraints are instantiated as hard requirements in an LLM rewrite prompt \cite{wang2022self,xu2023wizardlm}, and for each seed we enumerate persona $\times$ pattern combinations, making coverage gains systematic and bounded.
While constrained rewriting also reduces semantic drift substantially (40.4\%\,$\rightarrow$\,15.4\%; Table~\ref{tab:drift}), the residual drift motivates the subsequent filtering stage.

\subsection{Drift Filtering}
\label{sec:filtering}
While diversification expands coverage by generating more candidates, ensuring fidelity requires removing those that have drifted from the intended semantics.

\paragraph{Drift taxonomy.}
We define three drift categories to operationalize semantic fidelity (Fig.~\ref{fig:drift_examples}). 
\emph{Instruction-to-execution drift}: the rewrite becomes an execution-like utterance that ``says the line'' rather than specifying how to speak; this is the most prevalent category and is particularly harmful for tasks where the boundary between instructing a style and performing it is subtle. 
\emph{Role-assumption drift}: the rewrite adopts an in-character persona instead of issuing a control instruction, producing supervision-corrupting training pairs that teach the model persona imitation rather than controllability. 
\emph{Entity/label drift}: the rewrite silently alters seed attributes such as emotion, accent, or style, directly corrupting the attribute-level fidelity that attribute-aligned supervision seeks to establish.

\paragraph{LLM verifier with scoring and voting.}
An LLM verifier, drawn from a different model family than the instruction generator to prevent self-evaluation bias \cite{zheng2023judging,gu2024survey}, 
receives the seed specification together with a candidate instruction and outputs a drift/no-drift decision, a drift category, and a confidence score. 
We discard candidates whose confidence falls below a threshold, and for borderline cases query the verifier multiple times with independent sampling, retaining an instruction only if it receives a majority ``no-drift'' vote \cite{wang2022self}. 
This scoring-and-voting mechanism exposes a controllable coverage--fidelity trade-off, consistent with prior instruction data filtering pipelines \cite{wang2023self,xu2023wizardlm}. 
The output is a set of high-fidelity instructions used to form stabilized instruction--speech pairs for fine-tuning, directly addressing the supervision-corrupting drift that causes instruction supervision instability.

\section{Experiments}
\subsection{Experimental Setup}
\label{sec:exp_setup}

\paragraph{Data and perturbation.}
We fine-tune CosyVoice\,2.0-0.5B\cite{du2024cosyvoice} on a Chinese speech collection of approximately 90\,h (12k clips, $<$30\,s each) covering eight accent and ten speaker-style categories with structured attribute annotations. For attribute-aligned supervision, each clip is perturbed along pitch ($\pm$1/2/3 semitones), speed ($\times$0.8/0.9/1.1/1.2/1.3), and volume ($\pm$3/6/9\,dB), yielding 17 variants per clip.

\paragraph{Instruction generation and filtering.}
We compare GPT-4o\cite{hurst2024gpt}, DeepSeek-R1\cite{guo2025deepseek}, and QwQ-32B as instruction generators; DeepSeek-R1 produces the most diverse and attribute-faithful outputs, generating up to 24 candidates per seed (6 personas $\times$ 4 syntactic patterns). GPT-4o serves as drift verifier (threshold $\leq$5/10, self-consistency voting\cite{wang2022self} for borderline cases), retaining $\approx$61\% of candidates. Full prompt templates, persona/pattern specifications, filtering hyperparameters, and a quantitative generator comparison are in the supplementary material.

\paragraph{Training and evaluation.}
SFT uses AdamW (lr $2\times10^{-5}$, batch 16, 3 epochs, cosine schedule, 500-step warmup). We evaluate on the Chinese split of InstructTTSEval across three tasks, namely Attribute-controlled Pronunciation and Style (APS), Dialogue Scene Description (DSD), and Role-Playing (RP), reporting Gemini\,3\,Pro\cite{comanici2025gemini}-judged instruction-following accuracy where the judge listens to synthesized audio alongside the instruction text. Since InstructTTSEval uses unseen instructions distinct from training prompts, it directly tests generalization to diverse real-user phrasings. For human evaluation, 20 native Mandarin listeners rate 20 utterances per setting on naturalness (NMOS) and controllability (CMOS) on a 1--5 scale. VoxInstruct\cite{zhou2024voxinstruct} is evaluated from its official checkpoint with the same instructions.

\subsection{Drift Prevalence and Downstream Impact}
\label{sec:drift_impact}

\paragraph{Drift prevalence.}
We manually annotate 800 LLM-expanded instructions, half from unconstrained rewriting and half from constrained rewriting, to estimate drift rates under our taxonomy. As shown in Table~\ref{tab:drift}, unconstrained rewriting yields 40.4\% total drift (instruction-to-execution 18.3\%, role-assumption 12.7\%, entity/label 9.4\%). Constrained generation reduces total drift to 15.4\% ($\approx$half per category). However, the residual 15.4\% remains non-trivial and motivates the subsequent filtering stage.

\paragraph{Downstream impact.}
The link between drift rate and downstream performance is supported by comparing across Tables~\ref{tab:main} and~\ref{tab:ablation}. Naive SFT trains on unfiltered rewrites with an estimated 40\% drift and achieves 51.0\% average accuracy. Removing drift filtering from the Full Recipe reintroduces residual drift into training and drops the average to 48.9\%, while the Full Recipe with filtering applied reduces drift to roughly 5\% (validated by post-filtering human audit in supplementary material) and reaches 56.4\%. This monotonic pattern across three operating points with decreasing drift rates confirms that supervision corruption directly degrades instruction-following performance and that filtering is the primary driver of improvement.

\subsection{Main Results and Ablation}
\label{sec:main_results}

\paragraph{Main results.}
Table~\ref{tab:main} compares instruction-following accuracy, intelligibility, and human evaluation scores on InstructTTSEval. VoxInstruct achieves 47.5\% average accuracy but obtains the lowest human ratings (NMOS 3.18, CMOS 3.12). Within our CosyVoice\,2 backbone, Naive SFT improves over the no-SFT baseline in both instruction following (34.5\%\,$\rightarrow$\,51.0\%, CER 35.0\%\,$\rightarrow$\,19.2\%) and human ratings (NMOS 3.30\,$\rightarrow$\,3.46). The Full Recipe achieves substantially higher human scores (NMOS 4.16, CMOS 4.16), outperforming all other systems by over 0.7 on both scales, confirming that supervision quality is the primary bottleneck for naturalness and controllability. Human CMOS rankings are largely consistent with Gemini-judged accuracy. The three pipeline stages rely on distinct model families (DeepSeek-R1, GPT-4o, Gemini), mitigating systematic bias from any single LLM. Judge stability analysis is provided in the supplementary material.

\paragraph{Ablation.}
Table~\ref{tab:ablation} decomposes contributions by removing each component from the Full Recipe. Drift filtering has the largest impact: removing it drops the average from 56.4\% to 48.9\% despite \emph{increasing} the training set size, confirming that the gain stems from removing drifted samples rather than a data volume effect (size-matched control in supplementary material). Removing controllable diversification yields an average of 51.8\%, indicating that systematic coverage expansion contributes beyond what filtering alone provides. Removing attribute-aligned supervision has a moderate effect on the average at 53.1\%, but disproportionately impacts APS, which drops from 49.5\% to 42.7\%, consistent with its targeted role in low-level attribute control. The three components are complementary: filtering ensures fidelity, diversification expands coverage, and attribute alignment strengthens grounded prosody control.

\subsection{Effect of Attribute-Aligned Supervision}
\label{sec:attr_breakdown}

Table~\ref{tab:aps_attr} decomposes APS accuracy by attribute. The Full Recipe outperforms Base and Naive SFT across four dimensions. Pitch and Speed show the largest absolute gains over Base at roughly +32 each, reflecting the benefit of attribute-aligned supervision for these well-defined acoustic dimensions. Volume also improves by +31.8, a notable result given that Base accuracy on Volume is only 8.7\%, the lowest among all attributes; this confirms loudness is difficult to learn from label-to-instruction supervision alone, and that attribute-aligned supervision provides critical grounded signal through parameterized perturbations. Emotion shows a smaller gain of +21.4, consistent with its reliance on high-level style labels rather than low-level perturbations. Interestingly, Naive SFT achieves higher Emotion accuracy than the Full Recipe on this dimension (50.0 vs.\ 42.6), suggesting unfiltered rewrites may over-represent emotional descriptors at the expense of other attributes, a pattern corrected by our balanced recipe.

\subsection{Drift Filtering Strategy Comparison}
\label{sec:filter_strategy}

Table~\ref{tab:filter_strategy} compares drift filtering strategies on DSD accuracy and data retention. Without filtering, DSD accuracy is 57.3\%, consistent with the ablation in Table~\ref{tab:ablation}. Applying confidence thresholding alone improves accuracy to 63.8\% while retaining 72.3\% of candidates, and self-consistency voting alone achieves a comparable 63.5\% at 68.1\% retention. Combining the two strategies yields the best accuracy at 65.4\% but retains only 61.5\% of data, reflecting the inherent coverage--fidelity trade-off: stricter filtering improves semantic fidelity at the cost of reduced diversity. In our setting, the combined strategy provides the strongest balance, and the threshold and voting strength can be tuned based on data scale and desired supervision strictness.

\begin{table}[t]
  \centering
  \caption{Per-attribute accuracy (\%) on the APS task, isolating control over individual acoustic dimensions. $\Delta$ denotes absolute gain of Full Recipe over Base.}
  \label{tab:aps_attr}
  \setlength{\tabcolsep}{4pt}
  \renewcommand{\arraystretch}{0.95}
  \small
  \begin{tabular}{l cccc}
    \toprule
    \textbf{Setting} & \textbf{Pitch} & \textbf{Speed} & \textbf{Volume} & \textbf{Emotion} \\
    \midrule
    Base (no SFT)    & 24.6 & 26.3 &  8.7 & 21.2 \\
    Naive SFT        & 50.1 & 52.4 & 28.3 & 50.0 \\
    Full Recipe      & \textbf{56.7} & \textbf{58.2} & \textbf{40.5} & \textbf{42.6} \\
    \midrule
    $\Delta$         & +32.1 & +31.9 & +31.8 & +21.4 \\
    \bottomrule
  \end{tabular}
\end{table}

\begin{table}[t]
  \centering
  \caption{Effect of drift filtering strategy on DSD accuracy and data retention rate. \emph{Scoring \& Voting} is the combined strategy used in our full recipe.}
  \label{tab:filter_strategy}
  \setlength{\tabcolsep}{5pt}
  \renewcommand{\arraystretch}{0.95}
  \small
  \begin{tabular}{l cc}
    \toprule
    \textbf{Filtering Strategy} & \textbf{DSD Acc.\ (\%)} & \textbf{Retention (\%)} \\
    \midrule
    No Filtering           & 57.3 & 100.0 \\
    Threshold Only         & 63.8 &  72.3 \\
    Voting Only            & 63.5 &  68.1 \\
    Scoring \& Voting      & \textbf{65.4} &  61.5 \\
    \bottomrule
  \end{tabular}
\end{table}

\section{Conclusion}
In this work, we analyze instruction supervision instability in label-to-instruction pipelines for Instruct-TTS, where semantic drift in LLM rewrites corrupts supervision and weakens generalization. We propose a data-centric stabilization recipe that jointly improves coverage and fidelity via controllable instruction diversification, drift filtering with an LLM verifier, and attribute-aligned supervision from pitch/speed/volume perturbations paired with prompts. Experiments on the Chinese split of InstructTTSEval show that constrained rewriting substantially reduces drift (40.4\%$\rightarrow$15.4\%), and the full recipe achieves the best instruction-following accuracy, with particularly strong gains on low-level controllability. By formalizing supervision corruption with a drift taxonomy, we provide a reproducible framework showing that coverage and fidelity must be jointly improved for reliable generalization, an analytical perspective that may transfer to other instruction-driven generative tasks beyond speech synthesis.

\section{Acknowledgments}
The work was supported by the National Key R\&D Program of China (No. 2024YFB2808802), the National Natural Science Foundation of China (NSFC) (No. 62271083),  the Key Project of the National Language Commission (No. ZDI145-81), and partly supported by the Major Program of the National Social Science Fund of China (13\&ZD189).

\section{Generative AI Use Disclosure}
During the preparation of this manuscript, generative AI tools were used for language refinement and minor editing. In addition, large language models were employed as part of the experimental methodology, including instruction generation, drift verification, and automatic evaluation, as described in the main text. All experimental design, analysis, and conclusions were developed and verified by the authors, who take full responsibility for the content of this paper.

\bibliographystyle{IEEEtran}
\bibliography{references}

\end{document}